\documentclass[twocolumn]{aastex62}
\usepackage{amssymb}
\usepackage{array,multirow}
\usepackage{comment}
\usepackage{enumerate}
\usepackage{bm}

\usepackage[flushleft]{threeparttable}

\newcommand{\degree}{\ifmmode {^{\circ}\ }\else$^{\circ}$\fi}

\newcommand{\Msun}{\ifmmode {M_{\odot}}\else${M_{\odot}}$\fi}
\newcommand{\Rsun}{\ifmmode {R_{\odot}}\else${R_{\odot}}$\fi}
\newcommand{\Porb}{\ifmmode {P_{\rm orb}}\else${P_{\rm orb}}$\fi}
\newcommand{\Pspin}{\ifmmode {P_{\rm spin}}\else${P_{\rm spin}}$\fi}
\newcommand{\Pdot}{\ifmmode {\dot{P}_{\rm spin}}\else${\dot{P}_{\rm spin}}$\fi}
\newcommand{\age}{\ifmmode {\tau_{\rm c}}\else${\tau_{\rm c}}$\fi}

\submitjournal{ApJL}

\shorttitle{DNS sub-populations}
\shortauthors{Andrews, J. J. \& Mandel, I.}

\begin{document}

\title{Double Neutron Star Populations and Formation Channels}

\author[0000-0001-5261-3923]{Jeff J. Andrews}
\affiliation{Niels Bohr Institute, University of Copenhagen, Blegdamsvej 17, 2100 Copenhagen, Denmark}
\email{jeff.andrews@nbi.ku.dk}

\author[0000-0002-6134-8946]{Ilya Mandel}
\affiliation{Monash Centre for Astrophysics, School of Physics and Astronomy, Monash University, Clayton, Victoria 3800, Australia}
\affiliation{OzGrav: The ARC Center of Excellence for Gravitational Wave Discovery}
\affiliation{Birmingham Institute for Gravitational Wave Astronomy and School of Physics and Astronomy, University of Birmingham,
Birmingham, B15 2TT, United Kingdom}
\email{ilya.mandel@monash.edu}

\begin{abstract}
In the past five years, the number of known double neutron stars (DNS) in the Milky Way has roughly doubled. We argue that the observed sample can be split into three distinct sub-populations based on their orbital characteristics: (i) short-period, low-eccentricity binaries; (ii) wide binaries; and (iii) short-period, high-eccentricity binaries. These sub-populations also exhibit distinct spin period and spindown rate properties. We focus on sub-population (iii), which contains the Hulse-Taylor binary. Contrary to previous analysis, we demonstrate that, if they are the product of isolated binary evolution, the $P_{\rm orb}$ and $e$ distribution of these systems requires that the second-born NSs must have been formed with small natal kicks ($\lesssim$25 km s$^{-1}$) and have pre-SN masses narrowly distributed around 3.2 \Msun. These constraints challenge binary evolution theory and further predict closely aligned spin and orbital axes, inconsistent with the Hulse-Taylor binary's measured spin-orbit misalignment angle of $\approx$20$^{\circ}$. Motivated by the similarity of these DNSs to B2127$+$11C, a DNS residing in the globular cluster M15, we argue that this sub-population is consistent with being formed in, and then ejected from, globular clusters. This scenario provides a pathway for the formation and merger of DNSs in stellar environments without recent star formation, as observed in the host galaxy population of short gamma ray bursts and the recent detection by LIGO of a merging DNS in an old stellar population.
\end{abstract}


\section{Introduction}
\label{sec:intro}

Shortly after the discovery of the first double neutron star (DNS), the Hulse-Taylor binary \citep[B1913$+$16;][]{hulse75}, theorists derived evolutionary sequences describing the formation of these exotic systems in isolated stellar binaries \citep{webbink75, flannery75}. Their evolution typically involves a complex interplay of multiple mass transfer phases (both stable and unstable), two supernovae, and the effects of mass loss, tides, and rotation \citep{srinivasan82, bhattacharya91, tauris06}. The canonical evolutionary sequence is described in detail in \citet[][and references therein]{tauris17}. As pulsar surveys began to find more NS binaries, including several DNSs, it was realized that, although extraordinary, the Hulse-Taylor binary was by no means unique.

\vspace{0.3in}

\begin{figure}
	\includegraphics[width=0.99\columnwidth]{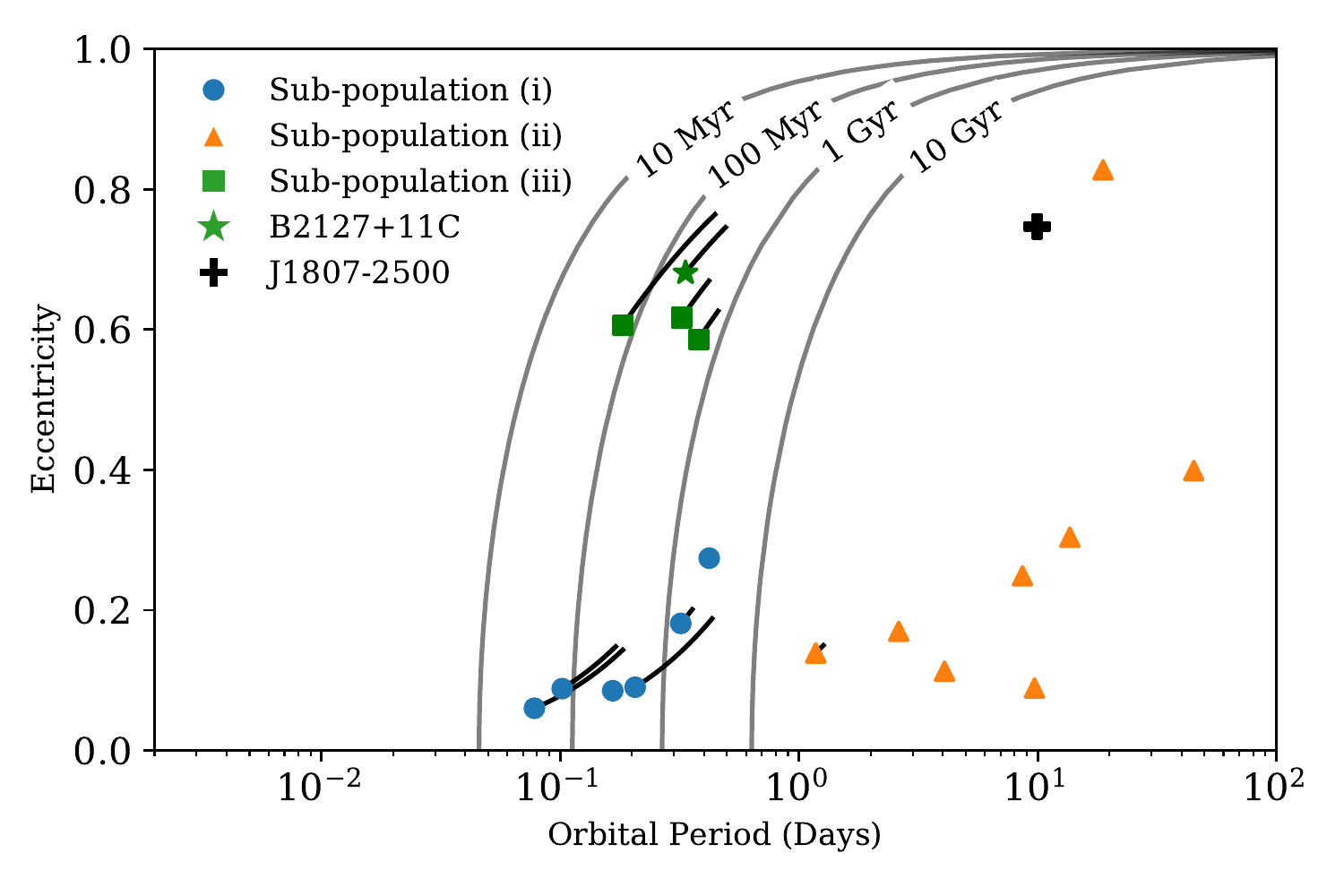}
    \caption{The currently known sample of DNSs in the field (separated into three sub-populations) and globular clusters (B2127$+$11C, J1807$-$2500). We overplot lines of constant merger time for $1.4~M_\odot$ components. Lines connected to each marker backtrack the prior orbital evolution due to general relativistic orbital decay using each system's characteristic age.} 
    \label{fig:DNS_pop}
\end{figure}

Stellar binary theory has progressed along with the expanding sample of Milky Way DNSs. Using binary population synthesis, a technique which evolves a population of high-mass binaries sampled from a distribution of initial conditions, studies generated synthetic DNS populations to uncover the properties of these systems, often with a focus on the rates of DNS mergers \citep{tutukov93, portegies_zwart98, belczynski99}. This technique is able to reproduce the broad characteristics of DNS populations \citep{portegies_zwart98, oslowski11, andrews15, shao18, vigna-gomez18, kruckow18}. The predictions of these studies were largely borne out with the recent detection of a merging DNS in gravitational waves by LIGO, GW170817, along with its electromagnetic counterpart \citep{ligo_detection}.

On the other hand, the detailed properties of DNS populations modelled with binary population synthesis do not always match the observed Galactic systems. Models predict too many systems with high eccentricity relative to observations \citep{ihm06, chruslinska17}, under-predict DNS masses \citep{vigna-gomez18}, and predict lower DNS merger rates than the point estimate based on the single gravitational-wave observation to date \citep[][although a single fortuitous detection is by no means implausible given the predicted range of merger rates]{belczynski18}.  

With recent observational efforts, there are now 17 known DNSs in the Milky Way field.
In Section \ref{sec:motivation} we demonstrate that the current sample clusters into three sub-populations, and we describe their general characteristics. In Sections \ref{sec:isolated} and \ref{sec:dynamical} we outline two separate formation scenarios, demonstrating that the short-period, high-eccentricity sub-population is best explained through dynamical formation. We end with a discussion of our results and conclusions in Section \ref{sec:conclusions}.

\begin{figure}
	\includegraphics[width=0.99\columnwidth]{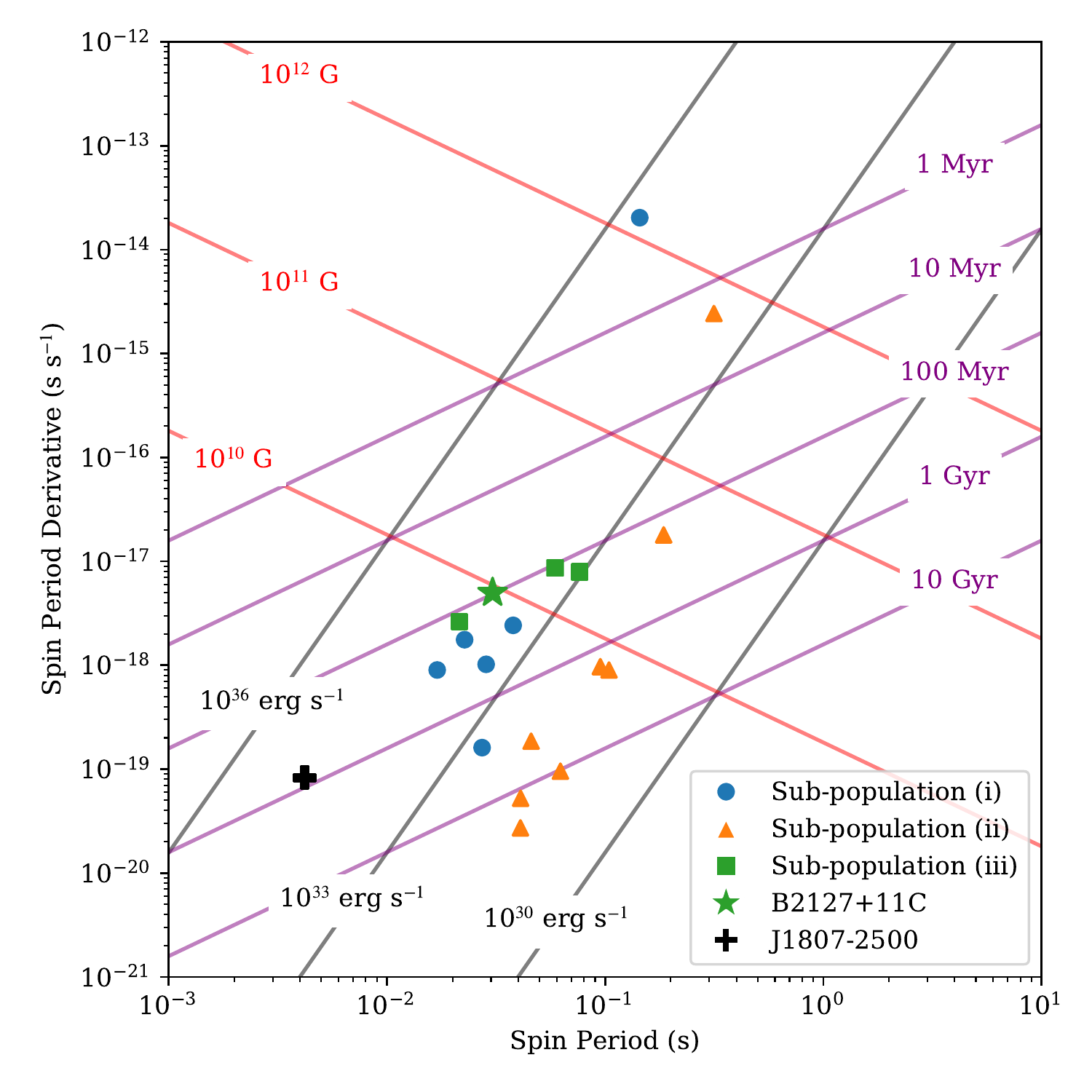}
    \caption{ The $P_{\rm spin}-\dot{P}_{\rm spin}$ diagram for the observed pulsars in DNSs. In addition to the clustering seen in Figure \ref{fig:DNS_pop}, the pulsars in the three DNS sub-populations we define also cluster in $P_{\rm spin}-\dot{P}_{\rm spin}$ space. For reference, we overplot lines of constant $B$-field (red), spin-down luminosity (black), and characteristic age (purple). To calculate these lines, we assume dipole radiation from a NS with a radius of 12 km, a mass of 1.4 \Msun, a moment of inertia of $(2/5) M R^2$, and an inclination angle of $\pi/2$ between the magnetic field and spin axes.} 
    \label{fig:DNS_p_pdot}
\end{figure}

\section{DNS Sub-populations}
\label{sec:motivation}

Figures \ref{fig:DNS_pop} and \ref{fig:DNS_p_pdot} show the orbital period ($\Porb$), eccentricity ($e$), spin period ($P_{\rm spin}$), and spin period derivative ($\dot{P}_{\rm spin}$) distributions of the current sample of 17 DNSs in the Milky Way field, along with the two known systems in globular clusters. The $\Porb$ -- $e$ distribution shows that systems with orbital periods $\Porb \lesssim1$ day exhibit a gap in eccentricity at $e \approx 0.4$. We apply an agglomerative clustering algorithm \citep{sklearn} to the $\Porb$ -- $e$ distribution of field DNSs to split the systems into three sub-populations. We highlight the clustering in Figures \ref{fig:DNS_pop} and \ref{fig:DNS_p_pdot} using separate colors and plot symbols: sub-population (i) contains binaries with small eccentricities that merge within a Hubble time; sub-population (ii) consists of widely separated systems that do not merge within a Hubble time; and sub-population (iii) includes binaries with orbital periods short enough to merge within a Hubble time but having eccentricities tightly clustered around 0.6. Clustering analysis formally associates the long-period, high-eccentricity system J1811$-$1736 with sub-population (iii), but we opt to group it with the other long-period systems. We list the sample of known DNSs, their properties, and their associated sub-populations in Table \ref{tab:DNS_properties}. 

Since they define the shape and size of the orbit, \Porb\ and $e$ are sensitive indicators of the prior dynamical evolution of the system; differences in this space are indicative of differences in SN dynamics and mass transfer sequences \citep{andrews19}. At the same time, pulsars in binary systems occupy different regions in $P_{\rm spin}$ -- $\dot{P}_{\rm spin}$ space depending principally on their accretion histories; pulsars that have accreted more mass during formation tend to achieve faster spin periods while simultaneously burying their magnetic fields (which reduces their $\dot{P}_{\rm spin}$), becoming ``recycled'' \citep{alpar82, radhakrishnan82, bhattacharya91}\footnote{The two outliers at the top right of Figure \ref{fig:DNS_p_pdot} are systems which show no recycling, suggesting the observed pulsar is the unrecycled, second-born NS; in these systems the primary may be unobserved due to an unlucky viewing angle.}. Taken together, the analogous clustering in both Figures \ref{fig:DNS_pop} and \ref{fig:DNS_p_pdot} \citep[cf.][]{tauris17} suggests that different formation channels are responsible for the three separate sub-populations.

\begin{table}[]
    \centering
      \begin{tabular}{lccccc}
  \hline
  System & \Porb & $e$ & \Pspin & \Pdot & \age \\
   & (days) & & (ms) & (10$^{-18}$) & (Myr) \\
  \hline
  \multicolumn{6}{c}{Sub-population (i)} \\
  \hline
  J0737$-$3039$^a$ & 0.102 & 0.088 & 22.7 & 1.76 & 204 \\
  B1534$+$12$^a$ & 0.421 & 0.274 & 37.9 & 2.42 & 248 \\
  J1756$-$2251$^a$ & 0.320 & 0.181 & 28.5 & 1.02 & 443 \\
  J1906$+$0746$^{a *}$ & 0.166 & 0.085 & 144.1 & 20300 & 0.1 \\
  J1913$+$1102$^a$ & 0.206 & 0.090 & 27.3 & 0.161 & 2687 \\
  J1946$+$2052$^b$ & 0.078 & 0.06 & 17.0 & 0.9 & 299 \\
  \hline
  \multicolumn{6}{c}{Sub-population (ii)} \\
  \hline
  J0453$+$1559$^a$ & 4.072 & 0.113 & 45.8 & 0.186 & 3901 \\
  J1411$+$2551$^c$ & 2.616 & 0.1699 & 62.4 & 0.0956 & 10342 \\
  J1518$+$4904$^a$ & 8.634 & 0.249 & 40.9 & 0.0272 & 23824 \\
  J1753$-$2240$^a$ & 13.638 & 0.304 & 95.1 & 0.970 & 1553 \\
  J1755$-$2550$^{a *}$ & 9.696 & 0.089 & 315.2 & 2430 & 2 \\
  J1811$-$1736$^a$ & 18.779 & 0.828 & 104.2 & 0.901 & 1832 \\
  J1829$+$2456$^a$ & 1.176 & 0.139 & 41.0 & 0.0525 & 12373 \\
  J1930$-$1852$^a$ & 45.060 & 0.399 & 185.5 & 18.0 & 163 \\
  \hline
  \multicolumn{6}{c}{Sub-population (iii)} \\
  \hline
  J0509$+$3801$^d$ & 0.380 & 0.586 & 76.5 & 7.93 & 153 \\
  J1757$-$1854$^e$ & 0.183 & 0.606 & 21.5 & 2.63 & 130 \\
  B1913$+$16$^a$ & 0.323 & 0.617 & 59.0 & 8.63 & 108 \\
  \hline
  \multicolumn{6}{c}{Globular Cluster} \\
  \hline
  B2127$+$11C$^a$ & 0.335 & 0.681 & 30.5 & 4.99 & 97 \\
  J1807$-$2500$^{a *}$ & 9.957 & 0.747 & 4.2 & 0.0823 & 805 \\
  \hline
  \end{tabular}
    \caption{
    The current sample of known and suspected DNSs in the field and in globular clusters. \\
    $^*$ Unconfirmed DNS. The DNS nature of several other systems is based principally on their non-zero eccentricities. In some cases, dynamical formation may allow NS-WD systems with non-zero eccentricities. \\
    $^a$ \citet{tauris17} and references therein. $^b$ \citet{stovall18}. $^c$ \citet{martinez17}. $^d$ \citet{lynch18}. $^e$ \citet{cameron18}.
    }
    \label{tab:DNS_properties}
\end{table}

\begin{figure*}
	\includegraphics[width=0.99\textwidth]{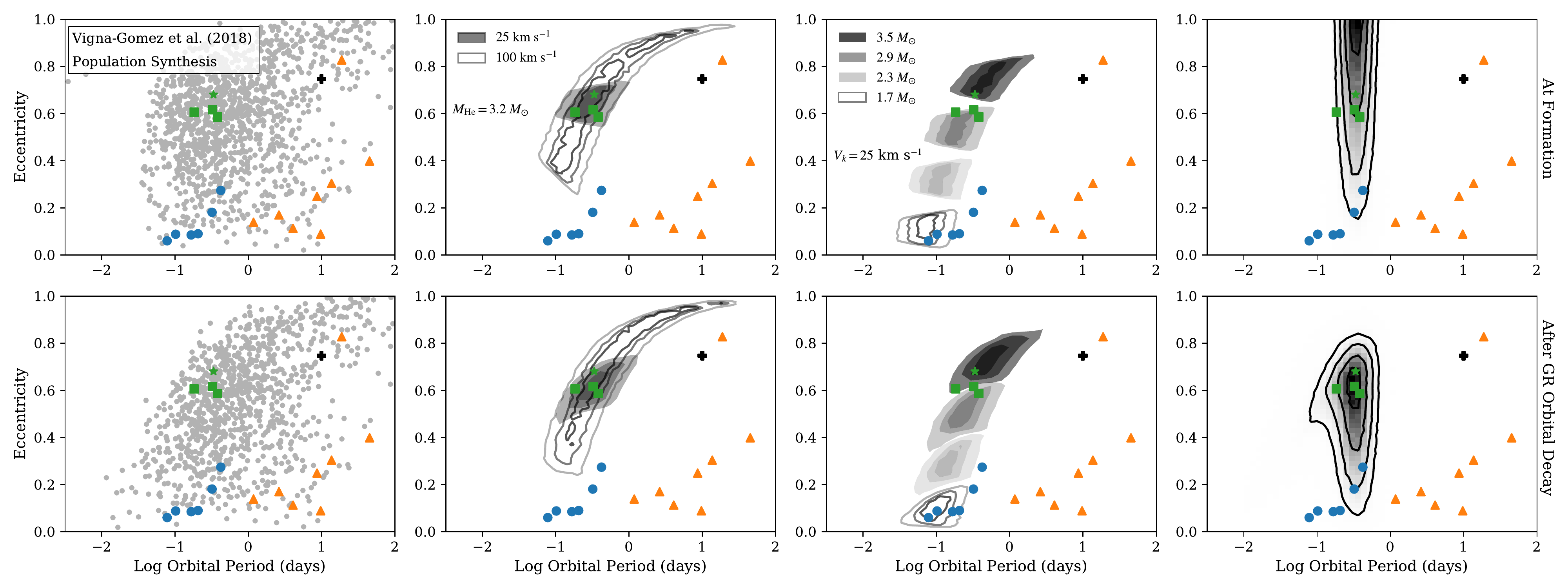}
    \caption{ In the left two columns of panels we compare the observed DNSs (plot symbols are identical to Figure \ref{fig:DNS_pop}) against the population synthesis predictions from 
    \citet{vigna-gomez18} both immediately after the second SN (top row) and after subsequent GR orbital decay (bottom row). The gap in observed systems with $0.3<e<0.6$ is unexpected from the models. The middle two columns of panels show that reproducing sub-population (iii) through isolated binary evolution requires SN kicks $\lesssim$25 km s$^{-1}$ from pre-SN He-stars with masses narrowly centered on $\approx$3.2 \Msun. Alternatively, the right column of panels shows a toy model for the dynamical formation of sub-population (iii). DNSs are generated with a thermal eccentricity distribution and a log-normal distribution in $a$ (top right panel). GR orbital decay causes the highest eccentricity systems to merge (bottom right panel), leaving a cluster of systems at eccentricities and orbital periods consistent with sub-population (iii). See Section \ref{sec:dynamical} for details.}
    \label{fig:analytic_post_SN}
\end{figure*}

\section{Isolated binary evolution}
\label{sec:isolated}

We first consider the formation of DNSs through isolated binary evolution. The \Porb\ and $e$ distribution of a population of DNSs resulting from isolated binary evolution is determined by the (distributions of) orbital separations before the second SN ($a$), pre-SN He-star masses ($M_{\rm He}$), SN kick velocities ($v_k$), and to a lesser extent, the systems' post-SN evolution driven by the emission of gravitational waves \citep{andrews19}. This simplicity is due to the fact that pre-SN orbits are expected to be circular\footnote{We expect that tides and mass transfer phases prior to the second SN are effective at circularizing orbits.} and SN kicks are assumed to be isotropically distributed in the orbital frame\footnote{Exploration of the possibility that NS kicks may be preferentially applied along a NS's spin axis, which may be aligned with the orbital axis due to past episodes of accretion, has not shown significant improvements in DNS models \citep[e.g.,][]{wang06, BrayEldridge:2018}.}.

Figure \ref{fig:DNS_pop} shows a clear gap in the observed eccentricity distribution ($0.3<e<0.6$) separating sub-populations (i) and (iii). In the top, left panel of Figure \ref{fig:analytic_post_SN}, we compare the known Galactic DNSs with the predictions for isolated binary evolution from the binary population synthesis model of \citet[][their fiducial model; gray points]{vigna-gomez18} for the population of DNSs immediately after the second SN. In the bottom left panel, we show the population after 100 Myr of orbital circularization and decay due to GR. While Figure \ref{fig:analytic_post_SN} shows that the population synthesis results can account for any individual DNS system, they predict a large number of systems within the observed eccentricity gap. Other binary population synthesis codes \citep[e.g.,][]{andrews15, kruckow18} predict similar distributions.

We focus on the second supernova during DNS formation in order to understand the physical reason behind these population synthesis results.  Using the equations for supernova dynamics from \citet{hills83}, we can calculate the post-SN distribution of \Porb\ and $e$, given pre-SN $a$, $M_{\rm He}$, and $v_{\rm k}$ \citep[see also][]{kalogera96,TaurisTakens:1998,andrews19}. For orbital modelling, we assume that all NSs have a mass of 1.4 \Msun. 

Broadly speaking, eccentricity in previously circular binaries can be produced by a combination of two types of kicks. The first are natal kicks imparted to the proto-NS upon collapse, responsible for the few hundred km s$^{-1}$ velocities of single pulsars in the Milky Way \citep{hobbs05}.  Larger natal kicks tend to produce systems with higher eccentricities, provided they stay bound. The second, so-called ``Blauuw'' kicks \citep{Blaauw:1961}, are due to instantaneous mass loss that is symmetric in the frame of the collapsing object, but asymmetric in the binary's center-of-mass frame. The eccentricity imparted to a circular binary by the Blaauw kick depends only on the fraction of the binary's mass lost during the SN, with the post-SN eccentricity given by the ratio of the mass lost to the remaining binary mass (the binary is disrupted if this ratio exceeds 1).  

By studying the evolution of massive post-common-envelope binaries comprised of a NS with a helium-main-sequence companion, \citet{tauris13, tauris15} showed that binary evolution theory predicts that systems with small Blaauw kicks ought to also receive small natal kicks; upon expansion after the He-main-sequence, the system will enter a second phase of mass transfer, which reduces the mass of the secondary's helium envelope to $\lesssim 0.5 M_\odot$.  When this ultra-stripped core explodes in a SN, there is little mass loss and limited asymmetry, yielding small natal and Blaauw kicks consistent with sub-population (i).   

On the other hand, to reproduce systems through this channel with eccentricities $e \sim 0.6$, such as those observed in sub-population (iii), the pre-SN binaries must have either large natal kicks, large Blaauw kicks, or both. In the top, middle-left panel of Figure \ref{fig:analytic_post_SN}, we show the post-SN orbital distributions assuming a log-normal pre-SN orbital separation distribution (centered on 1 \Rsun, with a standard deviation of 0.2 in log-space), a 3.2 \Msun\ pre-SN He-star, and a kick velocity of either 25 km s$^{-1}$ (gray contours) or 125 km s$^{-1}$ (open contours). Comparison between the two sets of contours demonstrates the well-known effect that large kick velocities produce post-SN distributions with large spreads in eccentricity. The bottom, middle-left panel of the same figure demonstrates that this conclusion is unaltered after GR circularization and orbital decay \citep[using the equations from][]{peters64} have taken place. 

Previous studies analyzing the detailed formation of the Hulse-Taylor binary through isolated binary evolution consistently suggested that it was formed with a kick velocity $\gtrsim$100 km s$^{-1}$ \citep{wang06, wong10, tauris17}. However our results in Figure \ref{fig:analytic_post_SN} indicate that systems formed with these large kicks ought to span a large eccentricity range, not form a cluster with $e\approx0.6$ like the three systems comprising sub-population (iii). Unless the observed sample is not representative of the overall Milky Way population (through either observational bias or small-number statistics), we conclude that sub-populations (i) and (iii) likely form through distinct evolutionary channels. We consider this our acting hypothesis throughout the remainder of this work.

The second column of panels in Figure \ref{fig:analytic_post_SN} shows how the tight clustering in sub-population (iii) can be reproduced with $v_k\approx$25 km s$^{-1}$ and $M_{\rm He}\approx 3.2 \Msun$. In the third column of panels in Figure \ref{fig:analytic_post_SN}, using the same distribution of pre-SN $a$ as was used in the second column of panels and a kick velocity of 25 km s$^{-1}$, we generate separate distributions of post-SN orbits for $M_{\rm He} =$ 1.7, 2.3, 2.9, and 3.5 \Msun. These distributions all produce eccentricities either too small or too large for the observed systems.

Although only three systems comprise sub-population (iii), the middle two columns in Figure \ref{fig:analytic_post_SN} show that -- assuming formation through isolated binary evolution -- in order to explain the observed clustering, the second-born NSs in sub-population (iii) systems must have kick velocities significantly less than 100 km s$^{-1}$ (larger kick velocities produce a DNS distribution spanning too wide a range of orbital separations and eccentricities) and a finely tuned distribution of $M_{\rm He}$ near 3.2 \Msun. We stress that this requirement arises from the need to statistically match the full population of observed DNSs in $P_{\rm orb}-e$ space, meaning both a high probability of producing observed systems {\it and} a low probability of producing systems in regions of the parameter space where there are no observations.

Our conclusions about the formation of sub-population (iii) challenge our understanding of isolated binary evolution. With a pre-SN mass of 3.2 \Msun, whatever mechanism \citep[e.g., ][]{janka17} causes the large natal kicks in single pulsars is unlikely to be suppressed. Furthermore, even if supernovae can produce low natal kicks from $3.2$ \Msun\ He-stars, some aspect of binary evolution must produce pre-SN orbits at a separation of $\sim$1 \Rsun\ with He-star masses of $\lesssim$2 \Msun\ (to explain sub-population (i)) and $\approx$3.2 \Msun\ (to explain sub-population (iii)), but no binaries with intermediate masses.

\section{Dynamical Formation}
\label{sec:dynamical}

Although DNS dynamical formation has been largely ignored in recent years \citep[although see][]{grindlay06,lee10}, there are two known DNSs that reside in globular clusters: B2127$+$11C in M15 \citep{anderson90} and J1807$-$2500 in NGC 6544 \citep{lynch12}. Figures \ref{fig:DNS_pop} and \ref{fig:DNS_p_pdot} show that one of these, B2127$+$11C, has \Porb, $e$, $P_{\rm spin}$, and $\dot{P}_{\rm spin}$ similar to the three field DNSs forming sub-population (iii). Either these similarities are coincidental or suggestive of a common evolutionary channel, in which case sub-population (iii) systems would have been formed in a dense stellar environment such as a globular cluster, then dynamically kicked into the Milky Way field\footnote{\citet{phinney91} propose this formation scenario for the Hulse-Taylor binary.}.

Provided the stellar cluster is largely devoid of black holes, which may decouple through the Spitzer instability and then be dynamically ejected \citep{Spitzer:1969} \citep[but see, e.g.,][]{Morscher:2013}, NSs will be the most massive objects in a cluster older than $\sim$ 5 Gyr. Subsequent mass segregation will cause the NSs to occupy the center of the cluster, where a series of capture, exchange, and fly-by interactions will form NS binaries. On average, these interactions harden binaries, which  sample a thermal eccentricity distribution, $p(e)=2e$ \citep{heggie75}.  This process continues until one of three things happens: (a) the binary runs out of objects to interact with (the interaction timescale becomes longer than the age of the Universe: $\tau_{\rm interaction}>\tau_{\rm Hubble}$), leaving the binary ``frozen'' in the cluster; (b) GR-driven evolution takes over as the dominant force for tightening the binary ($\tau_{\rm GW} < \tau_{\rm interaction}$), in which case a DNS will merge within the cluster; or (c) the binary is ejected from the cluster because the momentum recoil kick from the last interaction exceeds the cluster's escape velocity $v_{\rm esc}$, in which case a dynamically formed DNSs will be thrown into the Milky Way field.
  
\begin{figure}
	\includegraphics[width=0.99\columnwidth]{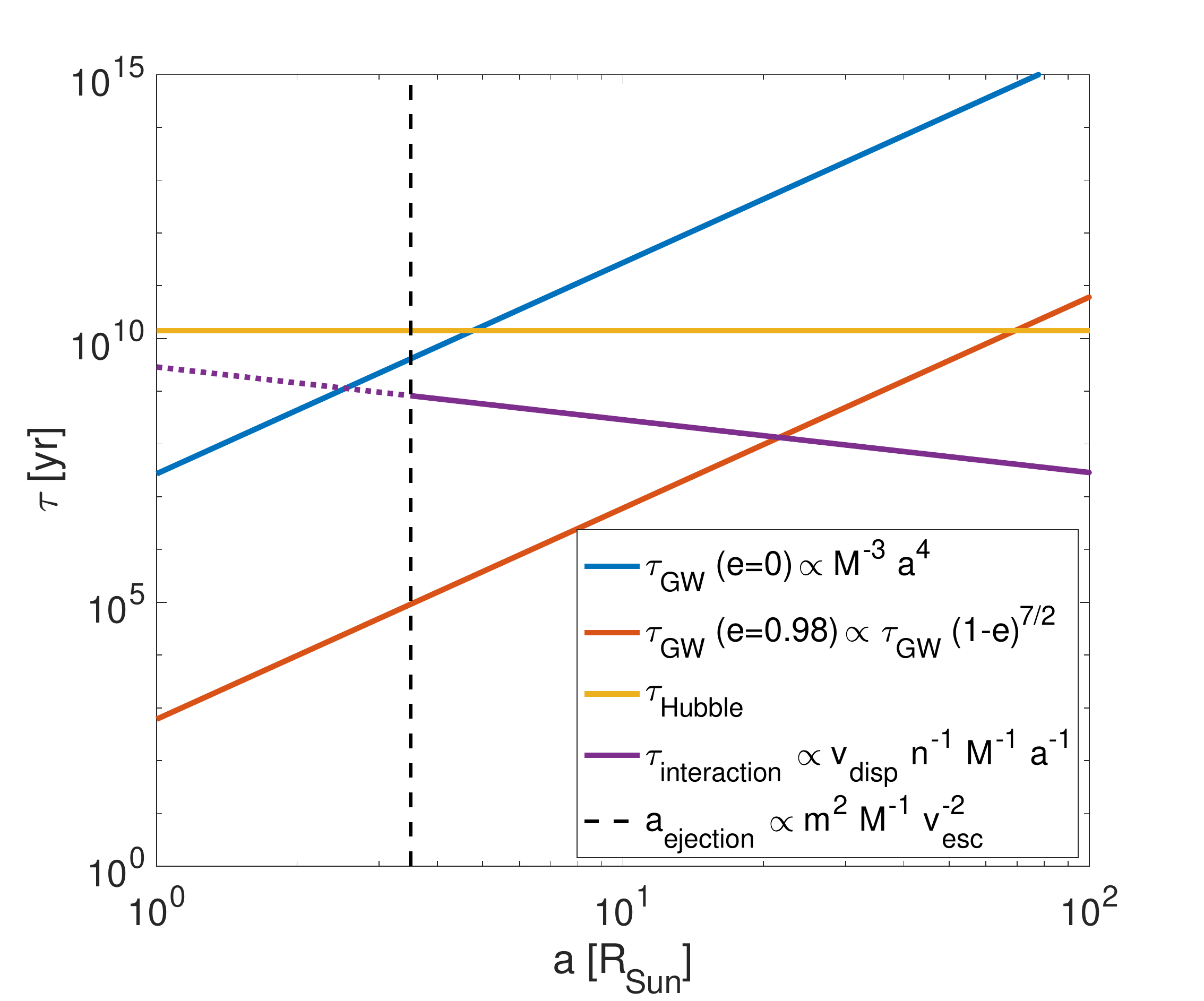}
    \caption{ Key timescales for the formation and ejection of DNSs from a globular cluster such as M15, as defined in Section \ref{sec:dynamical}.  The plot is made for a cluster with $n=2.5\times10^7$ pc$^{-3}$, $v_{\rm disp}=10$ km s$^{-1}$, $v_{\rm esc}=50$ km s$^{-1}$, and $M=m=1.4 M_\odot$.  The dashed vertical line illustrates the orbital separation of binaries that will typically be ejected from the cluster by a $2+1$ interaction.}
    \label{fig:timescales}
\end{figure}

These key timescales and their scalings with cluster properties are illustrated in Figure \ref{fig:timescales}. We adopt fiducial parameters taken from dynamical models of the expanded core of M15, the host of B2127$+$11C \citep{Dull:1997}: a central density $n=2.5\times10^7\ \textrm{pc}^{-3}$, a velocity dispersion $v_{\rm disp}=10$ km s$^{-1}$, and an escape velocity $v_{\rm esc}=50$ km s$^{-1}$, and assume an average interloper mass $m=1.4$ \Msun. As in our isolated binary models, we assume a NS mass of 1.4 \Msun. Comparison of the various timescales in Figure \ref{fig:timescales} shows that dynamical interactions will drive a DNS to smaller orbital separations, until it is ejected from the cluster with an average separation $a_{\rm ejection} \approx 3.5 \Rsun$ (vertical dashed line).

The ejection of such tight DNSs can only occur in clusters that have sufficiently high escape velocities and interaction timescales shorter than the age of the Universe. This likely requires core-collapsed  clusters, such as M15, as somewhat less dense and less massive globular clusters are likely to form DNSs at wider separations (J1807$-$2500 is such a system), if at all.

Dynamically formed binaries are likely to have high eccentricities \citep{heggie75}; however, since $\tau_{\rm GW}\sim(1-e^2)^{7/2}$, the highest eccentricity systems will evolve and merge first. This is illustrated in the right panels of Figure \ref{fig:analytic_post_SN}, where we assume a log-normal separation distribution for ejected binaries ($\mu=0.45$, $\sigma=0.1$ in log-space in units of R$_{\odot}$) and a thermal eccentricity distribution. We then give every system a random birth time uniformly chosen in the last 100 Myr (the characteristic ages of the sub-population (iii) systems are all between 100 and 150 Myr). The contours in the bottom right panel of Figure \ref{fig:analytic_post_SN} show the distribution of these systems in $\Porb$-$e$ space after GR orbital circularization and decay. The close match between the contours and the sub-population (iii) DNSs suggests that dynamical evolution can in principle yield tight clustering in \Porb -- $e$ space.

\section{Discussion \& Conclusions}
\label{sec:conclusions}

While some of the physics involved in binary evolution is complex and uncertain, the last two stages of DNS formation -- the dynamical effect of the supernova on the orbit and the subsequent orbital decay due to gravitational radiation -- involve well-understood physical processes. In this work, using the 17 known DNSs in the Milky Way field, we focus on those last two stages of DNS evolution and explore the conditions of the pre-SN DNS progenitors. Based on the observed systems' \Porb, $e$, $P_{\rm spin}$, and $\dot{P}_{\rm spin}$, we find that the distribution of systems may split into as many as three sub-populations. We focus on sub-population (iii) systems, characterized by large eccentricities and small orbital separations. If formed through isolated binary evolution, we find that sub-population (iii) would require relatively massive He-star progenitors $M_{\rm He}\approx$3.2 \Msun, receiving small SN natal kicks ($v_k\approx 25$ km s$^{-1}$). Larger kick velocities cannot reproduce the tight cluster of DNSs in $P_{\rm orb}-e$ space, while even modestly altered He-star masses predict DNSs with either smaller or larger eccentricities. 

Using additional information from the measured spin-orbit misalignment angle \citep[$\approx$$20^{\circ}$;][]{kramer98} and the position and velocity of the system in the Galaxy (using a dispersion measure-derived distance of 8.3$\pm$1.4 kpc)\footnote{\citet{deller18} recently published a VLBI parallax measurement placing the system at a distance of 4.1 kpc; this new distance halves both the Hulse-Taylor binary's height off the Galactic Plane, and its peculiar velocity.}, previous works on DNS formation have consistently argued that the Hulse-Taylor binary (a member of sub-population (iii)) requires a natal kick velocity of several hundred km s$^{-1}$ \citep[e.g.,][]{wong10, tauris17}. In particular, the spin-orbit misalignment angle of $\approx$20$^{\circ}$ is inconsistent with a low SN kick velocity. This inconsistency reinforces our analysis in Section \ref{sec:isolated}, which suggests that conventional binary evolution theory cannot simultaneously explain the low kicks and high He-star masses that we find best reproduces sub-population (iii) in $P_{\rm orb}$ -- $e$ space. 

We emphasize that previous binary population synthesis studies find that the Hulse-Taylor binary is at the short-period edge of a steeply declining distribution in $P_{\rm orb}-e$ space \citep{andrews15, vigna-gomez18, kruckow18}. Furthermore, none of these studies predict a gap in the eccentricity distribution of DNSs between 0.3 and 0.6. Either future observations will fill in the gap in short-period DNSs for $0.3<e<0.6$, or an alternative formation scenario is required. 

Based on their similarities with B2127$+$11C, a known DNS found in a globular cluster, we argue that sub-population (iii) DNSs in the field may have been dynamically formed in a dense stellar environment, such as a globular cluster, and kicked out into the field. We find that the birth cluster should be characterised by a high central density of at least a few million objects per pc$^3$ and an escape velocity $\gtrsim 50$ km s$^{-1}$ -- a combination best achieved at the centers of collapsed-core globular clusters such as M15. 

Detailed simulations will need to address several challenges to this scenario. Black holes, which would otherwise readily substitute into DNS binaries, must be predominantly ejected from the cluster before the present epoch. Additionally, the pulsar had to be recycled, then substituted into a DNS binary, and finally ejected from the cluster within the $<$150 Myr characteristic age of sub-population (iii) systems\footnote{Note, however, that characteristic ages may be a poor proxy for the true age of recycled pulsars, especially those coming from globular clusters, where the pulsar may have undergone several accretion phases.}, which would seemingly require a very high interaction rate \citep[cf.][]{grindlay06}; though the last two steps could be combined if a pulsar in a tight low-mass X-ray binary interacted with a NS. Furthermore, the three sub-population (iii) binaries are all found within a few degrees of the Galactic Plane, which, taken at face value, appears to be inconsistent with the orbits of globular clusters in the Milky Way; however given that pulsar searches tend to narrowly focus on the Galactic Plane, even systems ejected from globular clusters may be more likely to be observed when crossing the plane. Finally, the young characteristic ages and relatively short merger times of sub-population (iii) binaries could indicate that implausibly large formation rates would be needed to explain their prevalence in the Milky Way field. Nevertheless, the existence of B2127$+$11C in the collapsed-core cluster M15 suggests that this formation channel indeed occurs, possibly providing a non-negligible contribution to the overall rate of DNS mergers. Since dynamical formation predicts no correlation between the spin and orbital axes, measurements of the spin-orbit misalignment angles for more DNSs in sub-population (iii) will help discern between different formation scenarios.

Regardless of their exact formation channel, the short merger times of sub-population (iii) DNSs, combined with their young ($<$150 Myr) characteristic ages, have profound implications. If they are formed through isolated binary evolution, the short delay times between formation and merger of sub-population (iii) systems potentially provide a route to the $r$-process enrichment of globular clusters and ultra-faint dwarf galaxies \citep{ji16a}. While isolated binary evolution also produces some DNSs with long delay times \citep{vigna-gomez18, kruckow18}, if sub-population (iii) systems are formed dynamically in old globular clusters many Gyr after a star formation episode, they will naturally explain the abundance of DNS mergers and short gamma-ray bursts in non-star forming galaxies \citep{grindlay06}, potentially including NGC 4993, the host galaxy of the recently detected DNS merger GW170817  \citep{ligo_detection}.

\section*{Acknowledgements}

We thank Kyle Kremer, Claire Ye, Carl Rodriguez, Michele Trenti, and Thomas Tauris for useful discussions. We additionally thank the referee for comments and suggestions which greatly improved the quality of the manuscript. J.J.A.\ acknowledges support by the Danish National Research Foundation (DNRF132).

\bibliographystyle{aasjournal}

% \bibliography{references} 

\begin{thebibliography}{}
\expandafter\ifx\csname natexlab\endcsname\relax\def\natexlab#1{#1}\fi
\providecommand{\url}[1]{\href{#1}{#1}}

\bibitem[{{Abbott} {et~al.}(2017){Abbott}, {Abbott}, {Abbott}, {Acernese},
  {Ackley}, {Adams}, {Adams}, {Addesso}, {Adhikari}, {Adya}, \&
  et~al.}]{ligo_detection}
{Abbott}, B.~P., {Abbott}, R., {Abbott}, T.~D., {et~al.} 2017, Physical Review
  Letters, 119, 161101

\bibitem[{{Alpar} {et~al.}(1982){Alpar}, {Cheng}, {Ruderman}, \&
  {Shaham}}]{alpar82}
{Alpar}, M.~A., {Cheng}, A.~F., {Ruderman}, M.~A., \& {Shaham}, J. 1982, \nat,
  300, 728

\bibitem[{{Anderson} {et~al.}(1990){Anderson}, {Gorham}, {Kulkarni}, {Prince},
  \& {Wolszczan}}]{anderson90}
{Anderson}, S.~B., {Gorham}, P.~W., {Kulkarni}, S.~R., {Prince}, T.~A., \&
  {Wolszczan}, A. 1990, \nat, 346, 42

\bibitem[{{Andrews} {et~al.}(2015){Andrews}, {Farr}, {Kalogera}, \&
  {Willems}}]{andrews15}
{Andrews}, J.~J., {Farr}, W.~M., {Kalogera}, V., \& {Willems}, B. 2015, \apj,
  801, 32

\bibitem[{{Andrews} \& {Zezas}(2019)}]{andrews19}
{Andrews}, J.~J., \& {Zezas}, A. 2019, arXiv e-prints, arXiv:1904.06137

\bibitem[{{Belczy{\'n}ski} \& {Bulik}(1999)}]{belczynski99}
{Belczy{\'n}ski}, K., \& {Bulik}, T. 1999, \aap, 346, 91

\bibitem[{{Belczynski} {et~al.}(2018){Belczynski}, {Askar}, {Arca-Sedda},
  {Chruslinska}, {Donnari}, {Giersz}, {Benacquista}, {Spurzem}, {Jin},
  {Wiktorowicz}, \& {Belloni}}]{belczynski18}
{Belczynski}, K., {Askar}, A., {Arca-Sedda}, M., {et~al.} 2018, \aap, 615, A91

\bibitem[{{Bhattacharya} \& {van den Heuvel}(1991)}]{bhattacharya91}
{Bhattacharya}, D., \& {van den Heuvel}, E.~P.~J. 1991, \physrep, 203, 1

\bibitem[{{Blaauw}(1961)}]{Blaauw:1961}
{Blaauw}, A. 1961, Bull.~Astron.~Inst.~Netherlands, 15, 265

\bibitem[{{Bray} \& {Eldridge}(2018)}]{BrayEldridge:2018}
{Bray}, J.~C., \& {Eldridge}, J.~J. 2018, \mnras, 480, 5657

\bibitem[{{Cameron} {et~al.}(2018){Cameron}, {Champion}, {Kramer}, {Bailes},
  {Barr}, {Bassa}, {Bhandari}, {Bhat}, {Burgay}, {Burke-Spolaor}, {Eatough},
  {Flynn}, {Freire}, {Jameson}, {Johnston}, {Karuppusamy}, {Keith}, {Levin},
  {Lorimer}, {Lyne}, {McLaughlin}, {Ng}, {Petroff}, {Possenti}, {Ridolfi},
  {Stappers}, {van Straten}, {Tauris}, {Tiburzi}, \& {Wex}}]{cameron18}
{Cameron}, A.~D., {Champion}, D.~J., {Kramer}, M., {et~al.} 2018, \mnras, 475,
  L57

\bibitem[{{Chruslinska} {et~al.}(2017){Chruslinska}, {Belczynski}, {Bulik}, \&
  {Gladysz}}]{chruslinska17}
{Chruslinska}, M., {Belczynski}, K., {Bulik}, T., \& {Gladysz}, W. 2017,
  \actaa, 67, 37

\bibitem[{{Deller} {et~al.}(2018){Deller}, {Weisberg}, {Nice}, \&
  {Chatterjee}}]{deller18}
{Deller}, A.~T., {Weisberg}, J.~M., {Nice}, D.~J., \& {Chatterjee}, S. 2018,
  \apj, 862, 139

\bibitem[{{Dull} {et~al.}(1997){Dull}, {Cohn}, {Lugger}, {Murphy}, {Seitzer},
  {Callanan}, {Rutten}, \& {Charles}}]{Dull:1997}
{Dull}, J.~D., {Cohn}, H.~N., {Lugger}, P.~M., {et~al.} 1997, \apj, 481, 267

\bibitem[{{Flannery} \& {van den Heuvel}(1975)}]{flannery75}
{Flannery}, B.~P., \& {van den Heuvel}, E.~P.~J. 1975, \aap, 39, 61

\bibitem[{{Grindlay} {et~al.}(2006){Grindlay}, {Portegies Zwart}, \&
  {McMillan}}]{grindlay06}
{Grindlay}, J., {Portegies Zwart}, S., \& {McMillan}, S. 2006, Nature Physics,
  2, 116

\bibitem[{{Heggie}(1975)}]{heggie75}
{Heggie}, D.~C. 1975, \mnras, 173, 729

\bibitem[{{Hills}(1983)}]{hills83}
{Hills}, J.~G. 1983, \apj, 267, 322

\bibitem[{{Hobbs} {et~al.}(2005){Hobbs}, {Lorimer}, {Lyne}, \&
  {Kramer}}]{hobbs05}
{Hobbs}, G., {Lorimer}, D.~R., {Lyne}, A.~G., \& {Kramer}, M. 2005, \mnras,
  360, 974

\bibitem[{{Hulse} \& {Taylor}(1975)}]{hulse75}
{Hulse}, R.~A., \& {Taylor}, J.~H. 1975, \apjl, 195, L51

\bibitem[{{Ihm} {et~al.}(2006){Ihm}, {Kalogera}, \& {Belczynski}}]{ihm06}
{Ihm}, C.~M., {Kalogera}, V., \& {Belczynski}, K. 2006, \apj, 652, 540

\bibitem[{{Janka}(2017)}]{janka17}
{Janka}, H.-T. 2017, \apj, 837, 84

\bibitem[{{Ji} {et~al.}(2016){Ji}, {Frebel}, {Chiti}, \& {Simon}}]{ji16a}
{Ji}, A.~P., {Frebel}, A., {Chiti}, A., \& {Simon}, J.~D. 2016, \nat, 531, 610

\bibitem[{{Kalogera}(1996)}]{kalogera96}
{Kalogera}, V. 1996, \apj, 471, 352

\bibitem[{{Kramer}(1998)}]{kramer98}
{Kramer}, M. 1998, \apj, 509, 856

\bibitem[{{Kruckow} {et~al.}(2018){Kruckow}, {Tauris}, {Langer}, {Kramer}, \&
  {Izzard}}]{kruckow18}
{Kruckow}, M.~U., {Tauris}, T.~M., {Langer}, N., {Kramer}, M., \& {Izzard},
  R.~G. 2018, ArXiv e-prints, arXiv:1801.05433

\bibitem[{{Lee} {et~al.}(2010){Lee}, {Ramirez-Ruiz}, \& {van de Ven}}]{lee10}
{Lee}, W.~H., {Ramirez-Ruiz}, E., \& {van de Ven}, G. 2010, \apj, 720, 953

\bibitem[{{Lynch} {et~al.}(2012){Lynch}, {Freire}, {Ransom}, \&
  {Jacoby}}]{lynch12}
{Lynch}, R.~S., {Freire}, P.~C.~C., {Ransom}, S.~M., \& {Jacoby}, B.~A. 2012,
  \apj, 745, 109

\bibitem[{{Lynch} {et~al.}(2018){Lynch}, {Swiggum}, {Kondratiev}, {Kaplan},
  {Stovall}, {Fonseca}, {Roberts}, {Levin}, {DeCesar}, {Cui}, {Cenko},
  {Gatkine}, {Archibald}, {Banaszak}, {Biwer}, {Boyles}, {Chawla}, {Dartez},
  {Day}, {Ford}, {Flanigan}, {Hessels}, {Hinojosa}, {Jenet}, {Karako-Argaman},
  {Kaspi}, {Leake}, {Lunsford}, {Martinez}, {Mata}, {McLaughlin}, {Noori},
  {Ransom}, {Rohr}, {Siemens}, {Spiewak}, {Stairs}, {van Leeuwen}, {Walker}, \&
  {Wells}}]{lynch18}
{Lynch}, R.~S., {Swiggum}, J.~K., {Kondratiev}, V.~I., {et~al.} 2018, \apj,
  859, 93

\bibitem[{{Martinez} {et~al.}(2017){Martinez}, {Stovall}, {Freire}, {Deneva},
  {Tauris}, {Ridolfi}, {Wex}, {Jenet}, {McLaughlin}, \& {Bagchi}}]{martinez17}
{Martinez}, J.~G., {Stovall}, K., {Freire}, P.~C.~C., {et~al.} 2017, \apjl,
  851, L29

\bibitem[{{Morscher} {et~al.}(2013){Morscher}, {Umbreit}, {Farr}, \&
  {Rasio}}]{Morscher:2013}
{Morscher}, M., {Umbreit}, S., {Farr}, W.~M., \& {Rasio}, F.~A. 2013, \apjl,
  763, L15

\bibitem[{{Os{\l}owski} {et~al.}(2011){Os{\l}owski}, {Bulik},
  {Gondek-Rosi{\'n}ska}, \& {Belczy{\'n}ski}}]{oslowski11}
{Os{\l}owski}, S., {Bulik}, T., {Gondek-Rosi{\'n}ska}, D., \& {Belczy{\'n}ski},
  K. 2011, \mnras, 413, 461

\bibitem[Pedregosa et al.(2011)]{sklearn} Pedregosa, F., et al.\ 2011, J.\ Mach.\ Learn.\ Res, 12, 2825

\bibitem[{{Peters}(1964)}]{peters64}
{Peters}, P.~C. 1964, Physical Review, 136, 1224

\bibitem[{{Phinney} \& {Sigurdsson}(1991)}]{phinney91}
{Phinney}, E.~S., \& {Sigurdsson}, S. 1991, \nat, 349, 220

\bibitem[{{Portegies Zwart} \& {Yungelson}(1998)}]{portegies_zwart98}
{Portegies Zwart}, S.~F., \& {Yungelson}, L.~R. 1998, \aap, 332, 173

\bibitem[{{Radhakrishnan} \& {Srinivasan}(1982)}]{radhakrishnan82}
{Radhakrishnan}, V., \& {Srinivasan}, G. 1982, Current Science, 51, 1096

\bibitem[{{Shao} \& {Li}(2018)}]{shao18}
{Shao}, Y., \& {Li}, X.-D. 2018, \apj, 867, 124

\bibitem[{{Spitzer}(1969)}]{Spitzer:1969}
{Spitzer}, Jr., L. 1969, \apjl, 158, L139

\bibitem[{{Srinivasan} \& {van den Heuvel}(1982)}]{srinivasan82}
{Srinivasan}, G., \& {van den Heuvel}, E.~P.~J. 1982, \aap, 108, 143

\bibitem[{{Stovall} {et~al.}(2018){Stovall}, {Freire}, {Chatterjee},
  {Demorest}, {Lorimer}, {McLaughlin}, {Pol}, {van Leeuwen}, {Wharton},
  {Allen}, {Boyce}, {Brazier}, {Caballero}, {Camilo}, {Camuccio}, {Cordes},
  {Crawford}, {Deneva}, {Ferdman}, {Hessels}, {Jenet}, {Kaspi}, {Knispel},
  {Lazarus}, {Lynch}, {Parent}, {Patel}, {Pleunis}, {Ransom}, {Scholz},
  {Seymour}, {Siemens}, {Stairs}, {Swiggum}, \& {Zhu}}]{stovall18}
{Stovall}, K., {Freire}, P.~C.~C., {Chatterjee}, S., {et~al.} 2018, \apjl, 854,
  L22

\bibitem[{{Tauris} {et~al.}(2013){Tauris}, {Langer}, {Moriya}, {Podsiadlowski},
  {Yoon}, \& {Blinnikov}}]{tauris13}
{Tauris}, T.~M., {Langer}, N., {Moriya}, T.~J., {et~al.} 2013, \apjl, 778, L23

\bibitem[{{Tauris} \& {Takens}(1998)}]{TaurisTakens:1998}
{Tauris}, T.~M., \& {Takens}, R.~J. 1998, \aap, 330, 1047

\bibitem[{{Tauris} \& {van den Heuvel}(2006)}]{tauris06}
{Tauris}, T.~M., \& {van den Heuvel}, E.~P.~J. 2006, {Formation and evolution
  of compact stellar X-ray sources}, ed. W.~H.~G. {Lewin} \& M.~{van der Klis},
  623--665

\bibitem[{Tauris} {et~al.}(2015)]{tauris15} Tauris, T.~M., Langer, N., \& Podsiadlowski, P.\ 2015, \mnras, 451, 2123 

\bibitem[{{Tauris} {et~al.}(2017){Tauris}, {Kramer}, {Freire}, {Wex}, {Janka},
  {Langer}, {Podsiadlowski}, {Bozzo}, {Chaty}, {Kruckow}, {van den Heuvel},
  {Antoniadis}, {Breton}, \& {Champion}}]{tauris17}
{Tauris}, T.~M., {Kramer}, M., {Freire}, P.~C.~C., {et~al.} 2017, \apj, 846,
  170

\bibitem[{{Tutukov} \& {Yungelson}(1993)}]{tutukov93}
{Tutukov}, A.~V., \& {Yungelson}, L.~R. 1993, \mnras, 260, 675

\bibitem[{{Vigna-G{\'o}mez} {et~al.}(2018){Vigna-G{\'o}mez}, {Neijssel},
  {Stevenson}, {Barrett}, {Belczynski}, {Justham}, {de Mink}, {M{\"u}ller},
  {Podsiadlowski}, {Renzo}, {Sz{\'e}csi}, \& {Mandel}}]{vigna-gomez18}
{Vigna-G{\'o}mez}, A., {Neijssel}, C.~J., {Stevenson}, S., {et~al.} 2018, ArXiv
  e-prints, arXiv:1805.07974

\bibitem[{{Wang} {et~al.}(2006){Wang}, {Lai}, \& {Han}}]{wang06}
{Wang}, C., {Lai}, D., \& {Han}, J.~L. 2006, \apj, 639, 1007

\bibitem[{{Webbink}(1975)}]{webbink75}
{Webbink}, R.~F. 1975, \aap, 41, 1

\bibitem[{{Wong} {et~al.}(2010){Wong}, {Willems}, \& {Kalogera}}]{wong10}
{Wong}, T.-W., {Willems}, B., \& {Kalogera}, V. 2010, \apj, 721, 1689

\end{thebibliography}

% \bsp	% typesetting comment
% \label{lastpage}
\end{document}